\documentclass[preprint,tightenlines,superscriptaddress,aps,showpacs]{revtex4}
\usepackage{epsfig}

\newcommand{\beq}{\begin{equation}}
\begin{document}
\preprint{ICN-UNAM-01-18}
\title{Quantum Collapse of a Small 
Dust Shell}
\author{A. Corichi}
\email{corichi@nuclecu.unam.mx}
\affiliation{Instituto de Ciencias Nucleares - FENOMEC\\
Universidad Nacional Aut\'onoma de M\'exico\\
A. Postal 70-543, M\'exico D.F. 04510, M\'exico.}
\author{G. Cruz} 
\affiliation{FENOMEC-IIMAS\\
 Universidad Nacional Aut\'onoma de M\'exico\\
A. Postal 20-726, M\'exico D.F. 04510, M\'exico.}
\author{A. Minzoni}
\affiliation{FENOMEC-IIMAS\\
 Universidad Nacional Aut\'onoma de M\'exico\\
A. Postal 20-726, M\'exico D.F. 04510, M\'exico.}
\author{P. Padilla}
\affiliation{FENOMEC-IIMAS\\
 Universidad Nacional Aut\'onoma de M\'exico\\
A. Postal 20-726, M\'exico D.F. 04510, M\'exico.}
\author{M. Rosenbaum}
\email{mrosen@nuclecu.unam.mx}
\affiliation{Instituto de Ciencias Nucleares - FENOMEC\\
Universidad Nacional Aut\'onoma de M\'exico\\
A. Postal 70-543, M\'exico D.F. 04510, M\'exico.}
\author{M.P. Ryan, Jr}
\affiliation{Instituto de Ciencias Nucleares - FENOMEC\\
Universidad Nacional Aut\'onoma de M\'exico\\
A. Postal 70-543, M\'exico D.F. 04510, M\'exico.}
\author{N.F. Smyth}  
\affiliation{Department of Mathematics and Statistics,
University of Edinburgh, The King's Building, Mayfield Road, Edinburgh,
Scotland, United Kingdom EH9 3JZ.}
\author{T. Vukasinac}
\affiliation{Departamento de Fisica, Universidad Michoacana de
San Nicolas de Hidalgo, Morelia, Michoacan, M\'exico.}

\date{\today}

\begin{abstract}
The full quantum mechanical collapse of a small relativistic dust shell is
studied analytically, asymptotically and numerically starting from the
exact finite dimensional classical reduced Hamiltonian recently derived
by H\'aj{\'\i}\v cek and Kucha\v r. The formulation of the quantum mechanics
encounters two problems. The first is the multivalued nature of the
Hamiltonian and the second is the construction of an appropriate self adjoint
momentum operator in the space of the shell motion which is confined to
a half line. The first problem is solved by identifying and neglecting orbits
of small action in order to obtain a single valued Hamiltonian. The second
problem is solved by introducing an appropriate lapse function. The resulting
quantum mechanics is then studied by means of analytical and numerical
techniques. We find that the region of total collapse has very small
probability. We also find that the solution concentrates around the classical
Schwarzschild radius. The present work obtains from first principles a quantum
mechanics for the shell and provides numerical solutions, whose behavior is
explained by a detailed WKB analysis for a wide class of collapsing
shells.
\end{abstract}


\maketitle

\section{Introduction}

It is sometimes believed that the study of symmetry-reduced models, or
mini-superspace models are of some importance in the understanding of
the Physics in the strong curvature limit. This would be true, for instance,
in the Big Bang or at the singularity formed by collapse of matter.
There are a number of reasons why one would like to study the problem of the
quantum collapse of a localized accumulation of matter.  For large
accumulations one might wish to understand the complete evolution of a
collapsing system.  A star, for example, might be expected to collapse
through its horizon to form a black hole.  The usual argument is that this
ends all possibility of knowing what happened to the original star.  However,
the process of Hawking radiation will eventually bleed energy from the black
hole, and, if quantum gravity is not taken into account, the original hole
will disappear, with the consequent loss of coherence of the original
system.

One suggestion that avoids this problem is to assume that any
curvature singularities inside the horizon are artefacts of assuming that
gravity is a purely classical system, and that quantum gravity could replace
these by some quantum state that shows no such singularity.  When Hawking
radiation has reduced the black hole to the point where quantum gravity
comes into play, one might expect to find some sort of quantum
remnant that might solve the problem of evaporation.  This problem,
even if we simplify the system drastically before quantization, is of
daunting complexity, involving successively, classical collapse, quantum
field theory on a curved background, and finally full quantum gravity
coupled to quantum matter.

A simpler problem is to consider
the pure quantum collapse of phenomenological matter uncoupled to any
field that allows radiation. While in general this involves turning off
natural couplings, there is one reduced system that allows us to avoid
radiative modes, at least classically.  This is the classical collapse of
a spherically symmetric accumulation of dust.  Birkhoff's theorem shows
that there is no gravitational radiation, and the dust matter has no
radiative component (even though one might expect the expulsion of
shells of matter).  An even more drastically reduced case is the collapse of
a spherical shell of dust, where the matter is constrained to lie on
a (dynamical) spherical shell, thus avoiding blow off problems.  This
system provides perhaps the simplest quantum collapse scenario, where
the effects of quantum gravity are uncontaminated by possible radiative
effects.  In principle one could use this model to study the collapse of
a large shell that would pass through its horizon while still classical
and evolve into some sort of quantum object near the classical curvature
singularity.  Without Hawking radiation this quantum object would always
be hidden from outside observers, but it might give us some insight into
the avoidance of singularities due to quantum effects.

Note, however, that if one treats the
Schwarzschild mass of the shell as a $c$-number (as we will below)
\footnote{recall that a $c$-number is a number independent of the phase
space point. A phase space dependent number (constant of the motion)
is called a $q$-number.}, the mass
will probably have a limit beyond which there is no valid solution to
the quantum problem.  This phenomenon would be the analogue of the limit
of the charge of the nucleus in the relativistic hydrogen atom problem
\cite{Lanlif}.  It is possible to make rough estimates of this
limit. Thus, if we consider a particle of mass $m$ orbiting in the 
gravitational
field of a shell of mass $M$, the particle will be quantum mechanical if the
radius of the shell is less that the gravitational Bohr radius of the
particle, $a^G_0 = \hbar^2/GMm^2$.  It also seems reasonable to demand that
$a^G_0$ be larger than the Schwarzschild radius of the shell for the
particle to be observable (note that this excludes the large-mass collapse
scenario defined above), i.e. $a^G_0 \geq GM/c^2$ (we ignore factors of order
one).  It is usual to argue that single-particle quantum mechanics is no
longer valid if $a^G_0$ is less that the Compton wavelength of the particle,
$\hbar/mc$.  The second and third of these conditions both give $\sqrt{Mm}/M_P
\leq 1$, where $M_P$ is the Planck mass.  For stellar mass shells and
elementary particles this bound is strongly violated, and elementary
particles are produced well inside the horizon.  For shell of roughly
asteroid size ($10^{17}$ gm), the  bound is saturated when the shell radius
is roughly the Compton wavelength of the electron.  We would expect production
of light particles well outside the Planck length, and our single-particle
quantum mechanics would break down.  However, since we exclude all fields,
we might hope that our formalism would hold until the mass of the particles
produced was equal to that of the shell.  In that case, the bound would
be saturated when the shell mass was equal to the Planck mass (and the
Schwarzschild radius equal to the Planck length).  Thus we expect, in
order for quantum mechanics to make sense, a limit on the mass of the shell
of roughly the Planck mass, similar to a sufficient condition for
self adjointness of the quantum operators given by
H\'aj{\'\i}\v cek \cite{haj} for one of the models we will discuss below.

Keeping in mind these caveats, we will investigate the quantum mechanical
collapse of a roughly Planck mass shell of dust, from a rather novel point of
view of assymptotics.  This problem has been considered by several authors
\cite{haj,karel1,karel2,fried,bijak}.
In all of these articles the Israel junction conditions \cite{israel} were
used to obtain a classical equation of motion for the curvature radius of
the collapsing shell relative to the proper time of an observer on
the shell. Once the equations were obtained, several possible
Hamiltonians leading to these equations of motion were proposed and
used to study the corresponding quantum mechanics. These Hamiltonians are
both local and non-local. A fairly complete study of a local Hamiltonian is
reported in \cite{karel2}. The main conclusion of the above cited works is
the existence of a discrete spectrum of bound states and a set of continuum
eigenstates.  It was also shown that the ground state hovers away from
the classical Schwarzschild radius \cite{karel2}, as might be expected from
our qualitative arguments.  For the non-local formulation similar results
have been obtained, but they are less complete \cite{haj,ber}. In this case
the spectrum again turns out to be mixed but there is no detailed analysis
given of the eigenfunctions.  All these quantizations have qualitatively
similar spectral properties.

It is also important to remark, however, that all the above work, excepting
\cite{karel1}, is not based on the Hamiltonian arising from the spherical
reduction of the action of general relativity. Thus the question of how
to perform and study the canonical quantization remains open.  An important
step in this direction is provided by the work of Kucha\v r and
H\'aj{\'\i}\v cek \cite{karel1,bijak}. In this paper the authors
obtain the classical equations of motion of the shell as the exact finite
dimensional reduction of the full relativistic Hamiltonian.

The purpose of the present work is to continue the program outlined in
\cite{karel1} and investigate the quantum mechanics of the Hamiltonian of
H\'aj{\'\i}\v cek, Bi\v c\'ak and Kucha\v r. (A variational procedure for
deriving such a Hamiltonian has also been considered recently in
\cite{gladush}). The first difficulty one must overcome arises from the
multivalued nature of the canonical Hamiltonian.  This leads us to adopt
a procedure of quantization which takes into account the relative
size of the action on the different sheets of the Hamiltonian. When only the
dominant action is taken into account, the problem can be formulated in terms
of a non-local Hamiltonian. In this formulation the factor ordering of
operators imposed by the requirement of dilational symmetry of the half line
(which is the space of values of the shell radius) forces a definite choice
of time, which is related to but not the same as the proper time on the shell.

Our procedure results in a non-local Schr\"odinger equation with vanishing
coefficients at the classical ``Schwarzschild radius'' of the shell.
Since our quantum mechanics relates to the internal parameters of
the shell, with no reference to the external geometry, we will use
the term ``Schwarzschild radius'' to mean the point where the curvature
radius of the shell is equal to a constant of motion that is the
Schwarzschild mass of the classical exterior geometry.  The wave function
that solves our non-local Schr\"odinger equation can be interpreted as the
probability density of finding the shell with a given curvature radius.
It is interesting to observe that non-local operators often appear in wave
problems \cite{whit}, while the vanishing of the main operator at the
``Schwarzschild radius'' is analogous to critical layer absorption in fluid
mechanics \cite{light}, so the equations considered here combine two features
which in classical fluid problems are separate.  For these reasons, modern
ideas of asymptotics \cite{whit,tim} can be used successfully to obtain
accurate information about the dynamics generated by our new nonlocal
Schr\"odinger equation.

The analysis of the solutions of the above mentioned quantum equation for the
shell radius is performed using a generalization to non-local problems of
WKB ideas \cite{tim}.  The use of WKB ideas in a similar context in
General Relativity has been explored in \cite{far,visser,ansoldi1,ansoldi2}.
In \cite{visser} expected values of interesting physical parameters are
calculated using the WKB ground state.  While in
\cite{far,ansoldi1,ansoldi2} tunelling amplitudes are computed for simple
non-local operators.  In the present work we perform a uniform WKB
analysis of the continuum and discrete spectrum in order to study the full
initial value problem for a collapsing shell.  We find that there is indeed
a phenomenon analogous to critical layer absorption, which decreases the
amplitude of the eigenfunctions and the scattering states beyond the
classical Schwarzschild radius.  Moreover, we give a complete qualitative
description of the dynamics.  The analysis is complemented by a detailed
numerical study (based on classical ideas in wave propagation \cite{forn}
adapted to the present situation) for a representative range of parameters.
The numerics give solutions that closely match the dynamics that was
predicted qualitatively by the asymptotic analysis, and the results show a 
decomposition of the initial condition into bound states (which concentrate
outside the ``Schwarzschild radius'') and a small probability current
leaking to infinity. We also show how the present model explains, due to the
presence of a critical-layer like behavior, the concentration of the solution
of an initial value problem, with initial values outside the
``Schwarzschild radius'', in the region outside this radius.

To complete our analysis, we compare our numerical solutions with those
obtained for a simpler non-local {\it ad hoc} model \cite{haj,fried} and
contrast their differences.  Finally we give arguments for how the present
formulation could be used to obtain local information (around the space-time
trajectory of the shell) about the space-time geometry.

The paper is organized as follows: In the second section we review the
variational formulation of the classical relativistic problem. The third
section studies the classical motion and solves the problem of canonically
quantizing the multivalued Hamiltonian. This section ends with a canonical
quantum formulation in terms of a non-local equation for the shell problem.
In the fourth section the Hamiltonian operator is analyzed both asymptotically
and numerically. The fifth section is devoted to the detailed numerical study
of the dynamics for representative values of the parameters and initial
conditions. The last section is dedicated to conclusions and suggestions for
further work. The non-standard details of matching for the WKB solution
are sketched in the Appendix.

\section{Formulation of the Problem}

In order to make our presentation as self contained as possible, as well as
to fix notation and conventions, we recall briefly some of the details of the
Kucha\v r-H\'aj{\'\i}\v cek \cite{karel1,bijak} construction of the canonical
dynamics of gravitational shells. The geometry of the shell is described by
\begin{equation}
ds^{2} = -\Lambda^{2} dt^{2} + R^{2}(t) d\Omega^{2}, \label{1}
\end{equation}
where the function $R(t)$ is the (area) radius of the shell and
$\frac{d\tau}{dt}= \Lambda$ with $\tau$ denoting the proper time on the shell.

The Lagrangian for the dust matter is taken to be of the form
\begin{equation}
{\mathcal L}_{\Sigma}^{M} = \int \frac{1}{2} M(\Lambda^{-1} {\dot T}^{2}
-\Lambda )dt, \label{2}
\end{equation}
where $M = 4\pi R^{2} \rho(t)$ is the total mass of the shell and
${\dot T}= -U_{t}$ is the temporal component of the dust velocity \cite
{karel1}.

The gravitational field Lagrangian  is expressed in terms of the variables
$V= \frac{{\dot R}}{\Lambda}$ -- the proper velocity -- and $F_{\pm} = 1 -
\frac{2M_{\pm}}{R}$, where the Schwarzschild mass $M_{\pm}$ can be different
on each side of the shell. For the special case of a collapsing shell $M_{-}
= 0$, so the space inside the shell is flat. Thus
\begin{equation}
{\mathcal L}_{\Sigma}^{G}= \int (M_{+} {\dot T}_{+} + [L])dt.\label{3}
\end{equation}
Here ${\dot T}_{+}$ is the Schwarzschild time which determines the embedding
of the shell in the space to the right of the shell. The function $L$ is
given by
\begin{equation}
L = \Lambda R[(F + V^{2})^{\frac{1}{2}} - V A(|F|^{-\frac{1}{2}} V)],\label{4}
\end{equation}
where
\begin{eqnarray}
A(\eta) &= & \sinh^{-1}\eta,\;\; {\rm for} \quad F>0\\
        &= & {\rm sgn}(\eta) |\cosh^{-1}|\eta||\;\;{\rm for} \quad F<0.\label{5}
\end{eqnarray}
The jump $[L]$ is obtained by using $F_{+} = 1 - \frac{2M}{R}$ on the right of
the shell and $F_{-}= 1$ on the left.

The total action is given by
\begin{equation}
{\mathcal L}= {\mathcal L}_{\Sigma}^{G}+ {\mathcal L}_{\Sigma}^{M}.\label{6}
\end{equation}

It is now possible to obtain the equations of motion. Variation with
respect to $M$ gives $\Lambda = {\dot T}$. Variation with respect to $T$
gives ${\dot M} = 0$. Variation with respect to $T_{+}$ gives $M_{+}=$
constant. Finally, variation with respect to $M_{+}$ gives ${\dot T}_{+} =
\Lambda F^{-1} \sqrt{F^{2} + V^{2}} $. The equation of motion is obtained
by varying with respect to ${\dot R}$. The energy condition which determines
the trajectories for a given proper mass $\cal M$  of the shell is
\begin{equation}
\sqrt{1-\frac{2M}{R} + V^{2}} -\sqrt{1 + V^{2}}= -\frac{\cal M}{R}.\label{7}
\end{equation}
The relative signs of the square roots are obtained from the requirement
that the shell motion generate a cup space-time \cite{karel1}.

This equation is taken as the starting point in the treatments in Refs.\
\cite{karel2,fried}. In these works there are several possible Hamiltonians
that produce the equations of motion (\ref{7}). However, The resulting
quantizations, as we have pointed out in the Introduction, are not
obtained from an exact dimensional reduction of the full general relativistic
Hamiltonian of the theory. This leaves open the
question (since the quantizations are not unitarily equivalent) of
which canonical quantization one should use, and how the properties of these
quantizations differ from those of {\it ad hoc} procedures.  In the first
place, one must obtain the classical Hamiltonian from the Lagrangian
(\ref{4}). This is done in \cite{karel1} by finding the appropriate canonical
momentum and then constructing, by a finite dimensional reduction, the
Hamiltonian.

The Legendre transformation is given in \cite{karel1}.  Note that what was
done there is equivalent to the full ADM procedure.  The gravitational
action leads to a Hamiltonian constraint instead of a Hamiltonian. By
choosing an internal time, we can solve the constraint for the momentum
conjugate to our time, which gives us a straightforward Hamiltonian
(the ADM Hamiltonian).  Classically, of course, this procedure is
exactly equivalent to solving the unconstrained system and applying
the constraint.  Quantum mechanically there is an enormous difference,
one procedure leading to an ADM Hamiltonian and the other to a
Wheeler-DeWitt equation.  In the ADM procedure, there is no guarantee that
the evolutions for different times and different Hamiltonians are unitarily
equivalent.  The Wheeler-DeWitt route leads to the well-known ``problem of
time'' in the solutions and difficulties in defining probabilities.  Since
we will use the ADM procedure and two different times in this article,
we will briefly discuss problems associated with our approach.  For the moment
we will just give the results from \cite{karel1} necessary for our
purposes.

First, the expression for the
canonical momentum $\tilde P= \frac{\partial {L}}{\partial V}$
must be inverted to obtain $V$ as a function of $\tilde P$.
It turns out that for
$0\leq R \leq 2M$ the inverse function is double valued (see Equations
(C33)and(C34) in \cite{karel1}). This poses a question of interpretation when
using $\tilde P$ instead of $V$.
In Ref.\ \cite{karel1} the total Hamiltonian constraint for the shell plus
gravity takes the form
\begin{equation}
H^G + {\cal M} = 0, \label{hcon}
\end{equation}
where the gravitational super-Hamiltonian $H^G$ is written in terms of
$\tilde P$ and $R$.  Since ${\cal M}$ is the conjugate variable to $\tau$,
the proper (or curvature) time on the shell, the natural time choice is
$\tau$, which gives a Hamiltonian ${\cal M} = H^G$.  In Ref.\ \cite{karel1}
there are several possible values for $H^G$ depending on whether $R$ is
greater than or less than $2M$.  We must
construct the appropriate single valued function from these for a viable
classical mechanics. To do this it is only necessary to
construct the two-sheeted surface joined
across the branch cut $0\leq R \leq 2M$ at $\tilde P = 0$.

\begin{figure}
\includegraphics[height=8cm,angle=0]{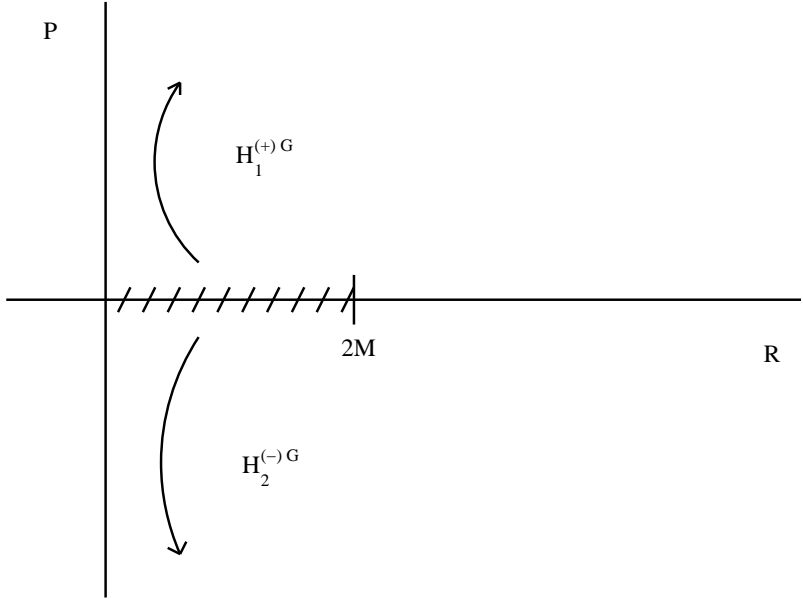}
\caption{\label{f:fig1}$R$--$P$ plane with cut in $0 \le R \le 2M$.}
\end{figure}

The patching condition across the cut (analytic continuation) is found
by taking into account the fact that $V$ does not change sign across the cut
Thus we have for $R\geq 2M$ that the Hamiltonian is given by
\begin{equation}
H^{G}_{1}= -\sqrt{2}R\left( 1- \frac{M}{R} - \sqrt{1- \frac{2M}{R}} \cosh
{\frac{\tilde P}{R}}\right )^{1/2}. \label{8}
\end{equation}
For $0\leq R \leq 2M$ and $\tilde P>0$ the Hamiltonian is given by
\begin{equation}
H^{(+)G}_{2}= -\sqrt{2}R\left (1- \frac{M}{R} - \sqrt{|1- \frac{2M}{R}|}
\;|\sinh {\frac{\tilde P}{R}}|\right )^{1/2}, \label{9}
\end{equation}
where the argument is positive. This is joined smoothly at $P= 0$ with
\begin{equation}
H^{(-)G}_{2}= -\sqrt{2}R\left (1- \frac{M}{R} + \sqrt{|1- \frac{2M}{R}|}
\;|\sinh {\frac{\tilde P}{R}}|\right )^{1/2}, \label{10}
\end{equation}
where the corresponding argument is positive and $\tilde P<0$. Likewise, the
other connection goes from $\tilde P<0$ to $\tilde P>0$ through the other
sheet. The points on the different sheets are then projected onto the
$R-\tilde P$ plane to obtain the classical orbit. This is illustrated in
Fig.\ \ref{f:fig1}.

It is to be noted that the double valuedness does not stem from the square
root but from the $\tilde P\rightarrow V$ transformation. Also note that in
the Legendre transformation the signs of $V$ and $\tilde P$ are not
necessarily the same. The velocity $V$ does not necessarily
change sign on a given orbit when
$\tilde P$ does (although in some cases it does),
due to the transformation. Thus the
Legendre transformation is only possible in terms of the double valued
Hamiltonian just described. As we shall see in the following Section, this
will have non-standard implications for the quantum theory based on this
Hamiltonian.

Notice that it is not necessary to interpret the geometrical variable
$\tau$ as time.  The constraint (\ref{hcon}) is simply a linear combination
of the geometrical variables $H^G$ and ${\cal M}$.

\section{Classical orbits, action and quantization}

In order to quantize we must first study the classical mechanics of the
problem and provide a suitable formulation for the multiple valued
Hamiltonian. For the classical mechanics we have to solve for the level lines
$H= E$. We anticipate these level lines to lie on two sheets. More
specifically, we take $R\geq 2M$. Thus $H^{G}= E$ gives for the level lines
the equation
\begin{equation}
\cosh{\frac{\tilde P}{R}} = f(E,R), \label{11}
\end{equation}
where
\begin{equation}
f(E,R)= \frac{(-E^{2}/2 + R^2 - MR)}{R^{3/2}\sqrt{R - 2M}}.\label{12}
\end{equation}

\begin{figure}
\centerline{
\hbox{\psfig{figure=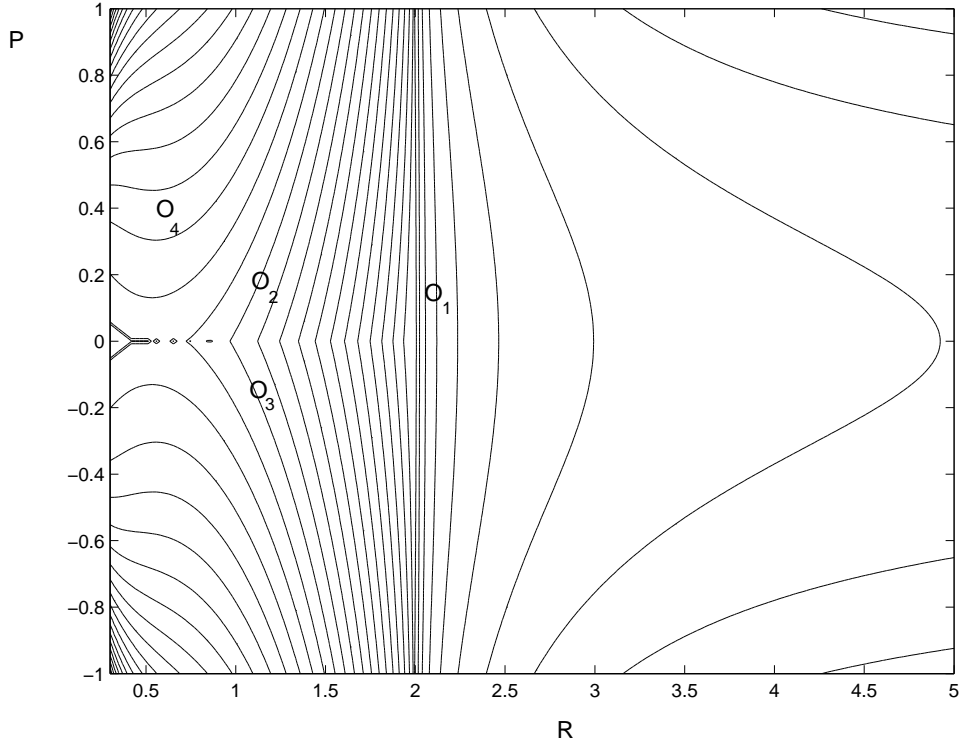,height=10cm,angle=0}}}
\bigskip
\caption{$(R,P)$ phase plane given by (\ref{11}), (\ref{12}), (\ref{14})
 and       (\ref{16}).}
\label{f:fig2}
\end{figure}

Provided $f(E,R)\geq 1$, Eq.\ (\ref{12}) can be
solved for $\tilde P$ real. There are
two solutions, one with $\tilde P>0$ and the other with
$\tilde P<0$. As $R\rightarrow
\infty$ there is a real solution only for $E^{2}\leq M^{2}$. For $M^{2}\leq
E^{2}\leq 2M^{2}$ the orbits do not
extend to infinity; they have a turning point at $R^*(E)$ given by
$f(E,R^*)= 1$, with
\begin{equation}
R^* (E)= \frac{ME^{2}+ \sqrt{M^{2}E^{4}+
E^{4}(E^{2}-M^{2})}}{2(E^{2}-M^{2})} = \frac{E^2(M + E)}{2(E^2 -
M^2)} > 0 \label{13}
\end{equation}
for $E^{2} \ge M^{2}$.  On the other hand, for $E^{2}\geq 2M^{2}$
there are no orbits which represent classical trajectories. For
these orbits, $\tilde P \rightarrow \infty$ as $R \rightarrow 2M$.
This part of the orbit is labeled $O_{1}$ in Fig.\ \ref{f:fig2}.

Let us now consider the situation when $R\leq 2M$. When $R$ approaches $2M$
from below, we have that the orbits are solutions of
\begin{equation}
|\sinh{\frac{\tilde P}{R}}|= f(E,R).\label{14}
\end{equation}
The solutions match as $|\tilde P|\rightarrow \infty$ since in that limit
$|\sinh{\frac{\tilde P}{R}}|\sim
\cosh{\frac{\tilde P}{R}}$. Thus the orbits can be
continued, and if we pick the collapsing orbits, i.e. $\tilde P<0$ they are
continued as solutions of (\ref{14}) up to $\tilde P= 0$.
We find that $\tilde P = 0$
at
\begin{equation}
R= {\tilde R}(E)= \frac{M+\sqrt{M^{2} +2E^2}}{2}, \label{15}
\end{equation}
obtained by solving $f[E,{\tilde R}(E)] = 0$. This orbit is labeled $O_{2}$ in
Fig.\ \ref{f:fig2}. Now, at this point the solution passes to the other sheet
and continues as a solution of
\begin{equation}
|\sinh{\frac{\tilde P}{R}}| = -f(E,R), \label{16}
\end{equation}
with $\tilde P>0$. This is labeled as $O_{3}$ in Fig.\ \ref{f:fig2}.  The
mirror image of the above description applies to the situation for expanding
orbits.  Finally for the case $E^{2} \leq M^2$ the situation is the same as
before but now the orbits extend as incoming or outgoing orbits as
$R\rightarrow \infty$.  These orbits are labeled in Fig.\ \ref{f:fig2} as
$O_{4}$.  This description reproduces the classical collapse as well as
expanding motion with turning points. This is to be contrasted with the phase
plane obtained in \cite{ansoldi2} where the local coordinate system used
gives only the $H^{G}_1$ portion of the Hamiltonian.  There the unbounded
orbits represent collapse up to a point and the bounded ones represent
expansion and ultimate collapse.  There is a separatrix dividing the phase of
these two types of motion. Also, in the quantum mechanics developed in the
above cited paper there is an exponentially small probability of total
collapse. In our analysis, in contrast, the orbits do not have a separatrix
and they all show ultimate collapse. We will show later that  the collapse
turns out to be localized around $R=2M$ via a phenomenon of critical layer,
the nature of which is unraveled by the Hamiltonian that we use. This
phenomenon is not captured by the local expansion used in \cite{far,ansoldi2}.
With this picture we turn now to the problem of quantization.

We will choose our internal time to be the proper time on the shell,
$\tau$, and $E$ is its conjugate variable.  We must now take $H^G$
to be an operator on an appropriate Hilbert space, and $R$ to be a
multiplication operator while $\cosh \tilde P/R$ and its functions
are to be realized as a pseudo differential operator on the Hilbert
space \cite{haj}.  However, in order to carry out this canonical quantization
we require a single-valued Hamiltonian, while, as already noted, the true
Hamiltonian of the shell given in Equations (\ref{9}) and (\ref{10})
is multi-valued.  This situation is new on both counts, and because of this we
need to introduce a proposal, which we believe to be new, to quantize the
system.  First we will consider possible definitions of the operator
$\cosh \tilde P/R$. H\'aj{\'\i}\v cek \cite{haj} proposes quantizing the shell
radius, which is restricted to the half line $R\geq 0$, by considering the
Hilbert space of functions $\Psi$ in $L^2 [0, \infty)$ with respect to the
measure $R^{-1} dR$.  The observables are taken to be $R$ and $\Pi = -iR
\frac{d}{dR}$.  The factor ordering is taken to make the operator
$\cosh \tilde P/R$ formally self-adjoint with $\tilde P = \Pi/R$ with
respect to the measure $R^{-1} dR$.  Since the hyperbolic cosine contains
only even powers of its argument, it is sufficient to take a factor ordering
that makes $(\Pi/R)^2$ formally self adjoint \cite{haj}.  In our case, due
to the $1/R$ factor, it is easy to check that this ordering is not possible.
This forces us, in order to preserve symmetries due to the dilational
symmetry of the half line, to redefine the $\cosh \tilde P/R$ operator
as simply $\cosh P$.  The operator $|\sinh P| = \sqrt{\cosh^2 P - 1}$ can
be shown to be self adjoint by means of spectral methods.

This quantization leads us to a new choice of time (lapse $\Lambda$) in
order to re-scale the momentum.  The freedom we have to choose $\Lambda$
allows us to construct a rescaled Hamiltonian and a canonical quantization
scheme. Thus if we now take $\Lambda = \mu/R$ in Eq.\ (\ref{1}), the
Lagrangian (\ref{4}) becomes
\begin{equation}
L = \mu \{(F + V^2)^{1/2} - VF(|F|^{-1/2} V)\} \label{17}
\end{equation}
Note that the resulting new canonical momentum obtained from
\begin{equation}
P = \frac{\partial L}{\partial V}
\end{equation}
does not contain the factor $1/R$.  Finally, Eq.\ (\ref{7}) is exactly
the same, except that now $V = \frac{\dot R R}{\mu}$, which allows us to
recover $\dot R$ from $V$.  The Hamiltonian with this new choice of lapse
is essentially the same as in Eqs.\ (\ref{8})--(\ref{10}),
except that $\tilde P/R$ is
replaced by $P$.  With our new time, we can obtain (as in Ref.\ \cite{haj})
$\cosh P$ as a formally self-adjoint operator.

The second difficulty we must address is the question of how to handle
the multi-valuedness of the Hamiltonian. By analogy to the path
integral formulation \cite{far}, we consider the classical action
$$\int^R_0 P dR$$ along an orbit, say $O_3$, $O_2$, or $O_1$.  Clearly the
action along the dashed, $O_3$, portion of the orbit is small compared to
the action along the $O_2$ and $O_1$ portions in solid line.  This occurs
because $R^* (E) >> \tilde R(E)$. An approach similar in spirit  was used in
\cite{far} where the action was approximated in the complex plane and then
the path integral was evaluated by a saddle point approximation. Here we
approximate the real action and obtain an approximate operator equation.
Also, since the classical Hamiltonian involves a square root, it is
necessary to define first the square and then take the square root using
spectral calculus. To this end we begin by defining $H$ as the square of
$H^{G}$ in the form
\begin{eqnarray}
H &=& 2 \{ (R^2 -MR) - R^{3/2} |R - 2M|^{1/2} \cosh P \}\;\;\;
R \geq 2M, \label{18a}\\
H &=& 2 \{ (R^2 - MR) - R^{3/2} |R - 2M|^{1/2} |\sinh P| \}\;\;
\;R \leq 2M, \label{18b}
\end{eqnarray}
once the small contribution of the $O_3$ portion of the orbit is neglected
This form of $H$ introduces classical turning points at $\tilde R(E)$ which
are not present in the original Hamiltonian.  The implications of this fact
will be discussed below.

We now turn to the quantization of (\ref{18a}-\ref{18b}).  As we have already
remarked, the operator $\cosh P$ is transformed into a self-adjoint operator
on functions $\Psi$ relative to the measure $dR/R$.  A unitary transformation
$R^{1/2} \Psi \rightarrow \Psi$ turns $\cosh P$ into a self-adjoint
operator in $L^2 (0, \infty)$ relative to the measure $dx$.  The
functions $\Psi$ must have all their even derivatives equal to zero at
$x = 0$. We thus have
\begin{eqnarray}
H\Psi(x) & = & 2\left\{ (x^2 - Mx)\Psi \right. \nonumber \\
 & & \mbox{} \left. - x^{3/2} |x -
2M|^{1/2} \frac{1}{2\pi} \int^{\infty}_{-\infty} \cosh k \:
e^{ikx} \hat
\Psi (k, t) dk \right\}, \nonumber \\
 & & \quad x \geq 2M,\label{19m} \\
 H\Psi(x) & = & 2\left\{ (x^2 - Mx)\Psi \right. \nonumber \\
  & & \mbox{} \left. - x^{3/2} |x - 2M|^{1/2}
\frac{1}{2\pi}\int^{\infty}_{-\infty} |\sinh k| \: e^{ikx} \hat
\Psi (k, t) dk \right\},  \nonumber \\
 & & \quad x \leq 2M, \label{20m}
\end{eqnarray}
where the Fourier transform $\hat{\Psi}$ of $\Psi$ is defined by
\begin{equation}
\hat \Psi (k, \tau) = \int^{\infty}_{-\infty} e^{-ikx} \Psi (x,
\tau) \: dx. \label{Schr}
\end{equation}
Now, since $H$ will be shown to be positive, the spectral calculus gives
the final equation in the form
\begin{equation}
i\frac{\partial u}{\partial{\tau}} = H^{1/2}u, \label{21n}
\end{equation}
and $u(x,0)=\Psi(x)$ is the given initial condition.  This procedure gives us
a well-defined equation for the quantum evolution of the radius of the shell.
If one prefers the Wheeler-DeWitt approach, one need only reinterpret
(\ref{21n}) as the equation that result from applying the Hamiltonian
constraint (\ref{hcon}) as a quantum equation, and think of $\tau$ as simply
a gravitational variable.  The solutions will be the same.

Before we solve the equation we need to make several remarks.  In the first
place, note that pseudo-differential equations like (\ref{21n}) arise in
classical wave propagation \cite {whit}, where the non-locality of the
main operator reflects the incompressibility of the fluid. The zero
coefficient at $x = 2M$ in front of the main operator also arises in the
study of wave propagation in stratified fluids in the presence of inviscid
critical layers \cite{light}.  However, in the case of critical layers the
main operator is local.  For these reasons we will use the same analytical
and numerical methods developed for those problems, with appropriate
modifications to take into account the presence of the vanishing coefficient
multiplying the main nonlocal operator in order to solve the problem posed
in (\ref{21n}).

One final remark is that we will treat $M$ as a $c$-number similar to the
charge on the electron in the hydrogen atom problem. Actually there is no
reason for this since $M$ is merely a constant of motion,
that is, a phase space function or
 $q$-number.   A simple analogue
might be a particle of mass $m$ moving in two dimensions, where in one
of the directions, $z$, the particle is free, and in the other, $y$, moves in
a harmonic oscillator potential.  The eigenfunctions have the form
$e^{ip_z z} \psi_n (y, p_z)$, where the $\psi_n$ are harmonic
oscillator eigenfunctions.  We may consider an eigenstate of $p_z$
and study the evolution of $\psi (y, t)$ with $p_z$ constant, but there is
no reason to do so.  One can just as well study more general functions of
the form $\int f(p_z) e^{ip_z z} e^{-i{{p_z^2}\over {2m}}t}
\psi (y, p_z) e^{-iE_n t} dp_z$.  In our case, we have taken an eigenfunction
of $M$, $\delta (M - M_0)$, but one should study more general situation.

Finally, we compare the solution of Equation (\ref{21n}) with another
model studied in \cite{haj,ber},
\begin{equation}
i\frac{\partial u}{\partial t}  = \frac{1}{2\pi}\int^{\infty}_{-\infty} \cosh k \: e^{ikx}
\hat u(k, t)dk - \frac{\beta u}{x}. \label{20}
\end{equation}
Note that further other models considered in the literature
\cite{fried} have similar structures with different symbols for
the nonlocal operator, or are similar to (\ref{20}) but with
functions of local operators.  The $\cosh k$ terms appear in all
models, since a proper time variable is used in their description.
The coordinate system chosen to cross the classical horizon
produces the nonlocal terms.  The potential term is common to all
the models.  However, the exact reduction leading to (\ref{21n})
has the new feature of having a vanishing term multiplying the
main operator.  We will discuss the implications of the
similarities and differences between the models in the following
sections.

\section{WKB solutions, spectra and the qualitative behavior of the
solutions}

As $x \rightarrow 2M$ the term in front of the nonlocal operator in (\ref{19m})
-(\ref{20m}) produces high frequency spatial oscillations.  Because of the
non-locality we look for an approximate solution in the WKB form of
(\ref{21n}). We take
\begin{equation}
u = A(x) e^{i\theta (x)} e^{-iE\tau}
\label{21}
\end{equation}
The phase $\theta$ satisfies the eikonal equation (\ref{11}) where $\tilde P/R$ is
replaced by $\theta^{\prime}$ (since $\tilde P/R$ is replaced by $P$ in the new
formulation), and the derivatives of $A$ are assumed to be small relative
to $\theta^{\prime}$.  The eikonal equation and amplitude equation are readily
obtained from the average Lagrangian \cite{whit,tim}
\begin{equation}
\int \int  [i(u^*\,\partial_{\tau}u  - u\,\partial_{\tau}u^* ) + 
u^{*} H u]\, dx dt.
\end{equation}
The eikonal equation takes the form
\begin{equation}
\cosh \theta^{\prime} = \frac{-E^{2}/2 + x^2 - Mx}{x^{3/2} |x - 2M|^{1/2}} =
f(E, x), \qquad x \geq 2M
\label{21a}
\end{equation}
\begin{equation}
|\sinh \theta^{\prime} | = \frac{-E^{2}/2 + x^2 - Mx}{x^{3/2}|x - 2M|^{1/2}}.
\qquad x \leq 2M
\label{21b}
\end{equation}

The first equation can be solved for real $\theta^{\prime}$ (which represents
an oscillatory solution as $x \rightarrow \infty$) provided $f(E, x) \geq
1$. This occurs for $x > 0$ provided $E^2 \leq M^2$.  On the other
hand, when $E^2 \geq M^2$ real solutions are only possible if $2M \leq x \leq
R^*(E) = \frac{E^{2}(M + E)}{2(E^{2} - M^2)}$.  Beyond $R^*$ the solution
becomes complex, and this region, $M^2 \leq E^{2} \leq 2M^2$, provides
candidates for the point spectrum.  On the other hand, the region $0 \leq
E^2 \leq M^2$ provides candidates for scattering states.  Note that $E \leq 0$
is not in the spectrum.  In the region $\tilde R(E) \leq x \leq 2M$ where
$\tilde R(E) = \frac{1}{2}(M + \sqrt{M^2 + 2E^{2}})$, Eq.\ (\ref{21b}) has
real solutions. Even though  $\theta^{\prime}$ goes to
infinity as $x \rightarrow 2M$ from above or below, $\theta^{\prime}$ matches,
since $\cosh \theta^{\prime} \sim |\sinh
\theta^{\prime}|$ as $|\theta^{\prime}| \rightarrow \infty$. For
$0 < x < \tilde R(E)$, $\theta^{\prime}$ is again complex.  This produces an
exponentially decaying solution as $x \rightarrow 0$.  This is a consequence
of the approximation we have used, which replaces a small action by zero in
the region $0 \leq x \leq \tilde R(E)$.  Also note that when $E^{2}<0$ there
are no solutions defined in the whole interval. This shows that $H$ is indeed
positive.

Given the phases, the amplitudes are determined from the average Lagrangian
by varying with respect to $\theta^{\prime}$.  Using
\begin{equation}
\int [H(\theta^{\prime}) A A^* - EA A^*]dx,
\end{equation}
we find that
\begin{equation}
\frac{\partial}{\partial x}\left (|A|^2 H(\theta^{\prime})\right ) = 0,
\end{equation}
which gives $|A|^2 = {\rm constant}/H(\theta^{\prime})$, where in each
region $\theta^{\prime}$ is chosen according to Eqs.\ (\ref{21a}) or
(\ref{21b}).

We now give the explicit expressions for the scattering states and for the
bound states.  We begin by finding the phases.  For $\tilde R(E) \leq x
\leq 2M$ we have
\begin{eqnarray}
\theta_1 (x) & = & \int^x_{\tilde R(E)} \ln \left\{
\sqrt{\frac{(\xi^2
- M\xi - E^{2}/2)^2}{\xi^3 (\xi - 2M)} + 1} \right. \nonumber \\
 & & \mbox{} \left. +
\frac{(\xi^2 - M\xi - E^{2}/2)} {\xi^{3/2} |\xi - 2M|^{1/2}}
\right \} \: d\xi, \label{22a}
\end{eqnarray}
while for $2M \leq x$ we have
\begin{eqnarray}
\theta_2 & = & \theta_1 (2M) + \int^x_{2M} \ln \left \{
\sqrt{\frac{(\xi^2 - M\xi - E^{2}/2)^2}{\xi^3 (\xi - 2M)} - 1}
\right. \nonumber \\
 & & \mbox{} \left. + \frac {(\xi^2 - M\xi - E^{2}/2)} { \xi^{3/2} (\xi -
2M)^{1/2}} \right \} \: d\xi. \label{22b}
\end{eqnarray}
Note that when $E^{2} \leq M^2$, $\theta_2 (x)$ is defined for all $x > 2M$.
On the other hand, when $E^{2} \geq M^2$, $\theta_2 (x)$ is real only up to
$x = R^*(E)$.  The amplitude associated with $\theta_1$ is given by
\begin{equation}
A_1 = \frac{1}{[(-E^{2}/2 + x^2 - Mx)^2 + x^3|x - 2M|]^{1/4}}, \qquad \tilde R
\leq x \leq 2M
\label{23a}
\end{equation}
while that associated with $\theta_2$ is
\begin{equation}
A_2 = \frac{1}{[(-E^{2}/2 + x^2 - Mx)^2 - x^3|x - 2M|]^{1/4}}. \qquad 2M \leq
x \label{23b}
\end{equation}
Note that when $x = 2M$, $A_1 = A_2$.  Also, when $0 \leq E^{2} \leq M^2$,
$A_2$ is always well defined for $ x \geq 2M$ and as $x \rightarrow \infty$
it is given by $A_2 (x) \sim \frac{1}{x^{1/2}}$.  On the other hand, when
$E^{2} \geq M^2$, as $x \rightarrow R^* (E)$, $A_2 \rightarrow \infty$.
This indicates, as usual \cite{whit,tim}, the presence of a transition region
which has to be included in order to construct the exponentially decaying
eigenfunctions of the point spectrum.  Finally, as $x \rightarrow \tilde R(E)$,
the amplitude $A_1$ is finite.  Due to the nonanaliticity of
$|\sinh \theta^{\prime}|$ at $\theta^{\prime} = 0$ when $x = \tilde R(E)$, the
behavior for $x < \tilde R(E)$ will be shown in the Appendix to be
algebraically small.  With these considerations we can now construct the
WKB functions which represent both scattering states and bound states.

We begin with the scattering states.  From the Appendix we have that the
solution for $x \sim \tilde R(E)$, but less that $\tilde R(E)$, matches to the
solution
\begin{equation}
A_1 (x) \sin \theta_1 (x).\label{aa1}
\end{equation}
In turn, this continues into
\begin{equation}
A_2 (x) \sin \theta_2 (x),\label{aa2}
\end{equation}
where $A_1$, $\theta_1$, $A_2$, $\theta_2$ are given by Eqs.\
(\ref{22a})-(\ref{23b}).  Thus the function $\varphi (x, E)$ defined by
Eqs.\ (\ref {aa1}) and (\ref {aa2}) is a solution of the
eigenvalue equation that is not square integrable and, as expected,
represents a perfectly reflected probability wave impinging from infinity.

The bound states occur for $E^{2} \geq M^2$.  In this case, as already noted
above, $A_2 \rightarrow\infty$ as $x \rightarrow R^* (E)$.  This is typical
turning point behavior, and the appropriate transition function will be shown
in the Appendix to be the Airy function.  Beyond $R^* (E)$ the phase is
complex and the solution decays exponentially.  The matching condition
between the oscillating and the Airy function gives the usual Bohr-Sommerfeld
condition for the eigenvalue $E$.  It takes the form
\begin{equation}
\theta_2 [R^* (E)]  = (n + \frac{1}{2}) \pi.
\label{24}
\end{equation}
This provides a transcendental equation for $E$.  It is clear from (\ref{24}),
since $\theta_2 [R^* (E)] \rightarrow \infty$ as $E^{2} \rightarrow M^2$
from above, that there are infinitely many eigenvalues accumulating at
$M^2$ from above.  An illustration of the W.K,B. eigenfunctions is given in
Fig.\ (3).  Notice that the oscillations are concentrated at the ``critical
layer,'' $x \sim 2M$.  It is also important to notice that the size of the
support of the oscillatory part of the eigenfunction which extends up to
$R^{*}(E)$ increases very rapidly even for low eigenvalues $(n = 3,4)$.

The eigenvalues of $H$ are the solutions to (\ref{24}) and the continuous
spectrum extends from zero to $M$. The analysis just presented shows also
explicitly that the eigenfunctions are localized around $R = 2M$ because of
the critical layer behavior.  Again, the existence of a point spectrum here
is to be contrasted with the results of \cite{ansoldi2} where the spectrum
is only continuous.  This effect was also found in other model Hamiltonians
\cite{karel2}, but no intrinsic explanation was possible. Finally, our
spectrum and eigenfunctions must be contrasted with those of other models.
In all models, including the present one, the spectrum is always mixed.
However, previous models have an unbounded continuous spectrum, and their
point spectrum has a ground state \cite{haj,karel2,ber}.  This situation just
reflects the fact that previous models were based on a choice of an
{\it ad hoc} Hamiltonian that reproduced the equation of motion.  Note,
nonetheless, that in principle the inverses of these {\it ad hoc}
Hamiltonians are equally legitimate choices, as they also reproduce the
classical equations of motion, but their spectrum will have a bounded
continuous part that is below the point spectrum.  Thus even if the quantum
problems are not unitarily equivalent, they have the same spectral properties
as the ones obtained from the finite dimensional reduced relativistic
Hamiltonian.  On the other hand, the form of the eigenfunctions in those models
was found to be similar to the ones for the hydrogen atom \cite{karel2,ber},
while for the Hamiltonian considered here the eigenfunctions and the
scattering states have a large number of oscillations concentrated around
$x = 2M$.

In the following section we consider the quantum dynamics generated by the
Schr\"odinger equation (\ref{21n}) and compare it with the dynamics of the
non-local Hamiltonian considered in \cite{karel2,fried,haj} and which leads to
\begin{equation}
i\frac{\partial u}{\partial t}= 
\frac{1}{2\pi}\int_{-\infty}^{\infty} \cosh k \:
e^{ikx}\;\; {\hat u} (k,t) dk - \frac{\beta u}{x}, \label{8m}
\end{equation}
However, before closing this section we would like to make some remarks
concerning the spectrum problem associated with (\ref{8m}) which show that
its eigenvalues and eigenfunctions behave as hydrogen-like. Thus
taking $u = v\exp {(-iEt)}$, the eigenvalue equation becomes
\begin{equation}
\hat{E}v= \frac{1}{2\pi}\int_{-\infty}^{\infty} \cosh k e^{ikx}\;\; {\hat
v} (k,t) dk - \frac{\beta v}{x}.\label{8p}
\end{equation}
The WKB eigenfunctions take the form
\begin{equation}
v= A e^{i\theta (x)}
\end{equation}
and the phase $\theta$ satisfies
\begin{equation}
\cosh {\theta^{\prime}} = E+\frac{\beta}{x}.\label{9m}
\end{equation}
Consequently for $E\geq 1$ there are always two real solutions for Eq.\
(\ref{9m}), and the case $E\geq 1$ is always part of the continuous
spectrum with eigenfunctions perfectly reflected at $x = 0$.
Alternatively, when $E < 1$ there are turning points for
\begin{equation}
E+ \frac{\beta}{x} = 1,
\end{equation}
which gives $x(E)= \frac{\beta}{(1-E)}$.
The quantization rule for the point spectrum is
\begin{equation}
\int_{0}^{x(E)} \cosh^{-1} 
\left( E+ \frac{\beta}{x}\right) dx = (n+ \frac{1}{2})\pi.
\label{10m}
\end{equation}
This equation shows that since $x(E) \rightarrow \infty$ as $E \rightarrow
1$, there are infinitely many eigenvalues accumulating from below at $E
= 1$. Also for $E = 0$ the integral is smaller than $\frac{\pi}{2}$. This
implies that the spectrum is bounded from below. We thus see that the
eigenvalues and eigenfunctions of (\ref{8p}) behave indeed as hydrogen-like
(see also \cite{ber}). With this information we shall next
consider numerical solutions of (\ref{8m}) and (\ref{21n}) and provide an
interpretation of the results.

\section{Quantum Dynamics and Numerical Solutions}

This section has three  parts. In the first one, we focus on the general
structure of the solutions of the quantum problem and show that, in
certain regime of interest, the quantum dynamics can be approximated by
a simpler Hamiltonian. In the second part, we describe the numerical scheme
used in the calculation. In the last part, we describe the numerical 
solutions for different initial conditions and compare them with the 
WKB solutions previously studied.

\subsection{Spectral properties and analytical results for the dynamics}

Let us begin by first remarking on the basic structure of the solutions
of the Schr\"odinger equations for the non-local Hamiltonians (\ref{19m})
and (\ref{20m}).  The solutions of the initial value problem take the form
\begin{equation}
u(x,\tau) = \sum_{n = 1}^{\infty} c_{n} \psi_n (x,E_{n}) e^{-iE_{n}\tau} +
\int_{0}^{M} c(E) \psi(x,E) e^{-iE\tau}\;dE.\label{11m}
\end{equation}
The $c_{n}$ and $c(E)$ are just the projections of the initial
condition on the discrete and continuous eigenfunctions
respectively.  The second term in (\ref{11m}) decays in time as
$\tau^{-1/2}$.  It is important to notice from equation
(\ref{22a}) that all the states in the continuous spectrum have
high wavenumbers in the vicinity of the critical layer.  It is
therefore expected that the radiation produced will be of short
wavelength in the vicinity of $2M$ and of long wavelength as $x
\to \infty$.  Since the eigenfunctions are strongly localized
between $R^{*}(E_{n})$ and ${\tilde R}(E_{n})$, the first term is
localized. Since $\psi\approx 0$ as $x \rightarrow 0$ the dynamics
will show small probability of total collapse.

We have shown that the discrete spectrum is concentrated around $E\sim M$,
while the continuous spectrum only contributes to the decaying part of the
solution. Thus the solution is dominated (in particular for large time which
is the relevant behavior for the physics of collapse) for values of the
spectral variable close to $M$. We can take advantage of this fact in
order to simplify considerably the numerical integration of (\ref{21n}).
To achieve this simplification use $\hat{H}^{1/2}=
\sqrt {(\hat{H}^{1/2})^2}$. Using
the spectral representation in terms of the eigenfunctions, we have
\begin{equation}
\hat{H}^{1/2} \varphi (x) = \sum_n \sqrt{E^{2}_{n}}\; c_{n} \psi_n (E_{n},x) +
\int_{0}^{M} \sqrt{E^{2}}\; c(E) \psi(E,x) dE.\label{100}
\end{equation}
Now, since the spectrum is concentrated around $E^{2}\sim M^{2}$, we have,
using the linear approximation around $M^{2}$,
\begin{equation}
\sqrt{E^{2}_{n}}= M + \frac{1}{2M} (E^{2}_{n} - M^{2}) + {\mathcal O}
(E^{2}_{n} - M^{2})^{2}.\label{101}
\end{equation}
This gives the approximation for $H^{1/2}$ in the form
\begin{equation}
\hat{H}^{1/2} \approx  \frac{M}{2} + \frac{1}{2M} H .\label{102}
\end{equation}
Thus, due to the nature of the spectrum, the equation
\begin{equation}
i\partial_{\tau}\psi = \hat{H}^{1/2} \psi  \label{schwar}
\end{equation}
is well approximated by
\begin{equation}
i\partial_{\tau}\psi = \left(\frac{M}{2} + \frac{1}{2M}\hat{H} 
\right)\psi.\label{103}
\end{equation}
With a change of variables $\psi = e^{-iM\tau/2} v(x,\tau/2M)$ the equation to
be solved numerically is
\begin{equation}
i\partial_{\bf t}v = \hat{H} v, \label{104}
\end{equation}
where ${\bf t} = \tau/2M$. This reflects the combination of two facts : first
that classically  $H^{1/2}$ and $H$ have the same orbits with rescaled
time, and second that since the discrete spectrum only runs over a small
energy range, the solution to (\ref{schwar}) is well approximated by the
solution to (\ref{104}).

\subsection{Numerical scheme and numerical stability}

We now outline the numerical scheme used for the solutions of the
equations. The equation to be solved is (\ref{104}). These
equation was solved numerically using a pseudo-spectral method
based on that of \cite{forn} for the Korteweg-de Vries equation.
This method involves finite Fourier transforms in the
$x$-direction, with the equation propagated forward in the
$t$-direction by fourth-order Runge-Kutta integration.  The finite
Fourier transforms are evaluated using fast Fourier transforms
(FFT).

The partial differential equation (\ref{104}) is numerically
solved in the interval $-L/2 \leq x \leq L/2$.  Let us denote the
Fourier coefficients of $v$ by $\hat{v}_n$ and let $\xi_n =
2n\pi/L$. Then the Fourier integral operators in (\ref{104}) can
be evaluated as
\begin{eqnarray}
\frac{1}{N} \sum_{n= -m}^{n= m} \cosh\xi_n \: \hat{v}_n e^{-i\xi_n
x}, & & \label{1M} \\
\frac{1}{N} \sum_{n= -m}^{n= m} |\sinh\xi_n| \: \hat{v}_n
e^{-i\xi_n x}, & & \label{1N}
\end{eqnarray}
respectively, where $2m+1$ terms are taken in the finite Fourier series.

Once the Fourier integrals have been evaluated, the partial
differential equation (\ref{104}) is then integrated forward in
time in Fourier space. This integration is carried out in Fourier
space due to its superior stability. Let us denote the right hand
side of (\ref{104}) by $Gv$. Then if $F$ denotes the finite
Fourier transform, we have
\begin{equation}
\frac{d}{dt} F(v)= -iF(Gv). \label{2N}
\end{equation}
The ordinary differential equation (\ref{2N}) is integrated using
fourth-order Runge-Kutta integration to find $F(v)$. The solution
$v$ is then determined by inversion using the FFT. The
pseudo-spectral method applied to the partial differential
equation (\ref{104}) has an inherent numerical stability problem
which will now be discussed.

The interface at $x= 2M$ is a boundary layer in which the solution
for $v$ rapidly oscillates. This rapid oscillation results in the
formation of high frequency modes in the finite Fourier series.
Then, due to the equivalence between small $x$ in physical space
and large $\xi_n$ in Fourier space, these high frequency modes
generate instabilities for $x\approx 0$ which then propagate into
the rest of the numerical domain. These high frequency modes are a
major problem for (\ref{104}) as the hyperbolic functions in the
Fourier integrals magnify their influence.  To reduce these
numerical instabilities, the numerical method as outlined above
was changed in two ways.  The first was that the governing
equation (\ref{104}) was changed to
\begin{eqnarray}
i \partial_{t}v & = & 2\left\{ (x^2 - Mx)v \right. \nonumber \\
 & & \mbox{} \left. - x^{3/2}
\left[ (x - 2M)^{2} + \epsilon ^{2} \right] ^{1/4} \frac{1}{2\pi}
\int^{\infty}_{-\infty} \cosh k \: e^{ikx} \hat v(k, t) \: dk
\right\}, \nonumber \\
 & & \qquad x \geq 2M, \label{e:epsg} \\
i \partial_{t}v & = & 2\left\{ (x^2 - Mx)v \right. \nonumber \\
 & & \mbox{} \left. - x^{3/2}
\left[ (x - 2M)^{2} + \epsilon ^{2} \right] ^{1/4}
\frac{1}{2\pi}\int^{\infty}_{-\infty} |\sinh k| \: e^{ikx} \hat
v(k, t) \: dk \right\},  \nonumber \\
 & & \qquad x \leq 2M.
\label{e:epsl}
\end{eqnarray}
The modification of the coefficient of the Fourier integral terms
in the equations was done to help resolve the boundary (critical)
layer.  The system (\ref{e:epsg}) and (\ref{e:epsl}) was solved as
previously explained in (\ref{2N}).  In the numerical solutions,
$\epsilon$ was taken to be $0.1$. The second modification was that
a low pass filter was applied when evaluating the integrals in
(\ref{e:epsg}) and (\ref{e:epsl}). This low pass filter is
equivalent to setting $\cosh \xi_n$ and $\sinh \xi_n$ to 0 for
large $|\xi_n|$. It is interesting to note that the concentration
of the spectrum at $E\sim M$ allowed us to control the instability
by using the very accurate linear approximation given in
(\ref{104}). In other problems it may turn out that the spectrum
is not concentrated in a small region, in which case one must find
the numerical square root of the matrix $FG$. This process must be
carried out in a manner consistent with the filtering, and this
procedure is currently being studied. The choice of the number of
modes used to obtain an accurate numerical solution was dictated
by comparison of the long time numerical solution with the WKB
solution as an initial condition with the same W.K.B solution.

\subsection{Numerical solutions}

\begin{figure}
\centerline{
\hbox{\psfig{figure=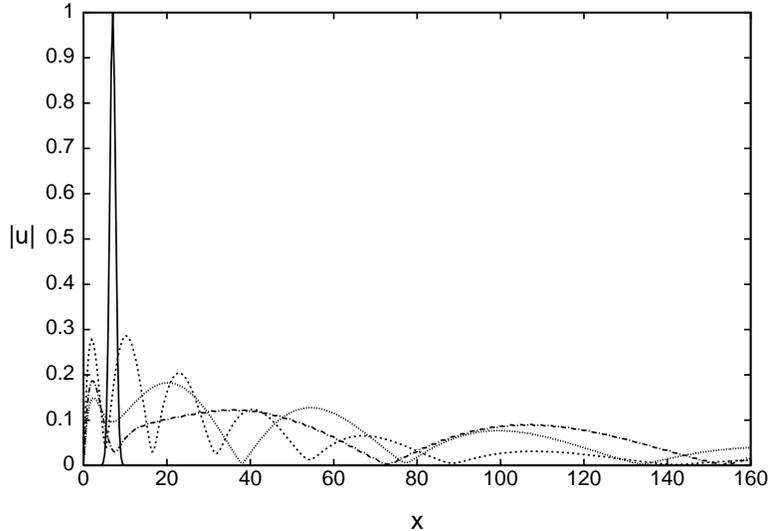,height=10cm,angle=270}}}
\bigskip
\caption{Numerical solution of model equation (\ref{20}) for initial condition
       (\ref{e:betaini}) with $x_{0} = 7$ and $\lambda = 1.0$.  Here
       $\beta = 0.5$.  $t = 0$: ~---~; $t=20$: ~--~--~--~; $t = 50$:
       $\cdots$; $t = 100$: ~-~--~-~--~-~.}
\label{f:beta}
\end{figure}

We begin by considering the simpler problem of solving Eq.\
(\ref{8m}) with a Gaussian as an initial condition.  In Fig.\
\ref{f:beta} we present the result for the integration of
(\ref{8m}) with the Gaussian initial condition
\begin{equation}
u(x,0) = e^{-(x - x_0)^{2}/\lambda},
\label{e:betaini}
\end{equation}
for the parameter values $\lambda=1.0$ and $x_{0}=7.0$, which was
reflected as an odd function about $x = 0$ to satisfy the boundary
conditions at $x = 0$.  This figure clearly shows the
decomposition of the initial condition into smooth normal modes
which are localized away from $x = 0$.  Notice that the modes are
smooth and resemble the hydrogen atom eigenfunctions, as can be
seen from equations (\ref{9m}) and (\ref{10m}) and the
representation in terms of Laguerre functions \cite{ber}.

\begin{figure}
\centerline{
\hbox{\psfig{figure=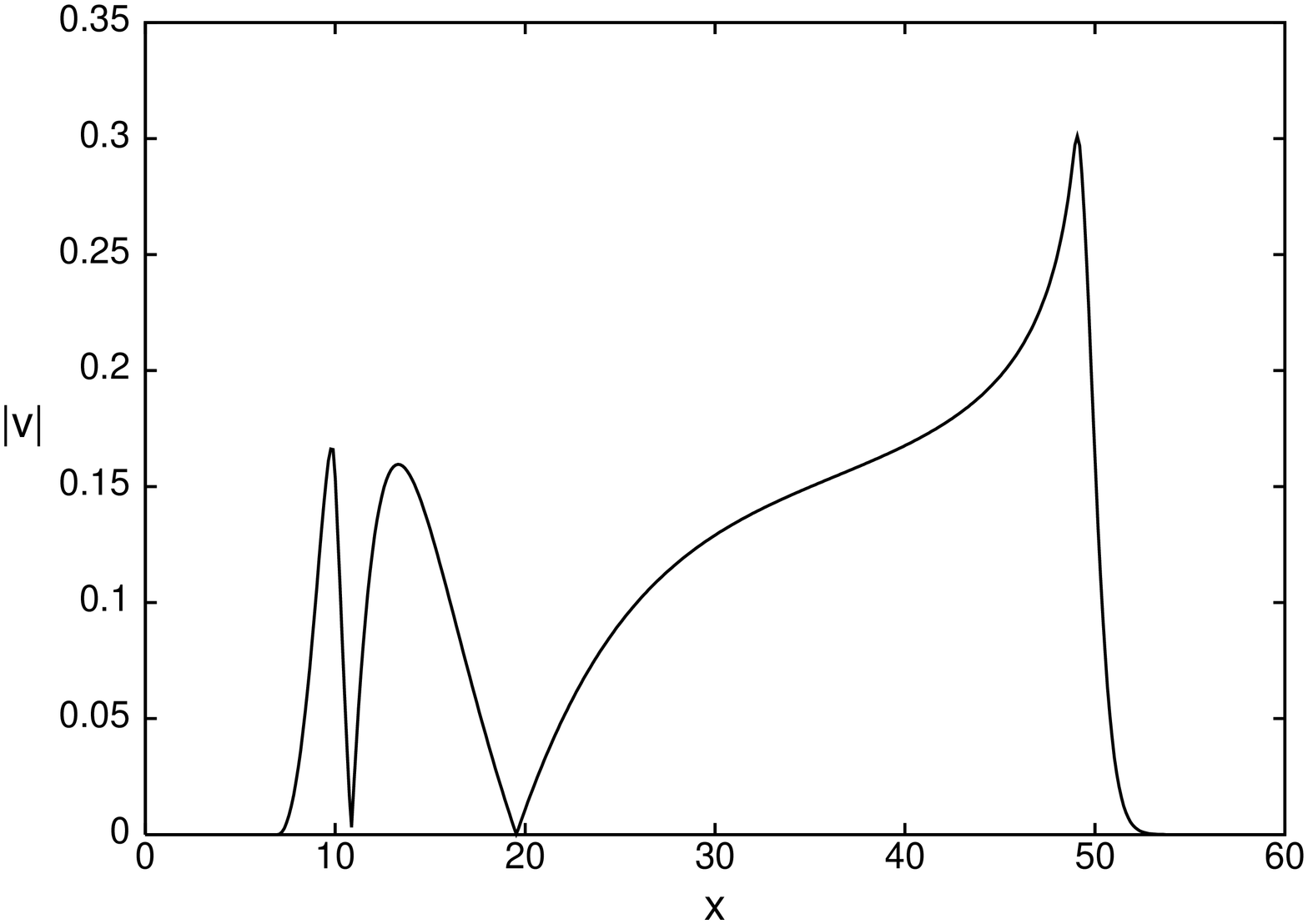,height=10cm,angle=270}}}
\bigskip
\caption{ WKB solution given by (\ref{aa1}) and (\ref{aa2}) with 
             $M = 10$ and $n = 3$.}
\label{f:wkb}

\centerline{
\hbox{\psfig{figure=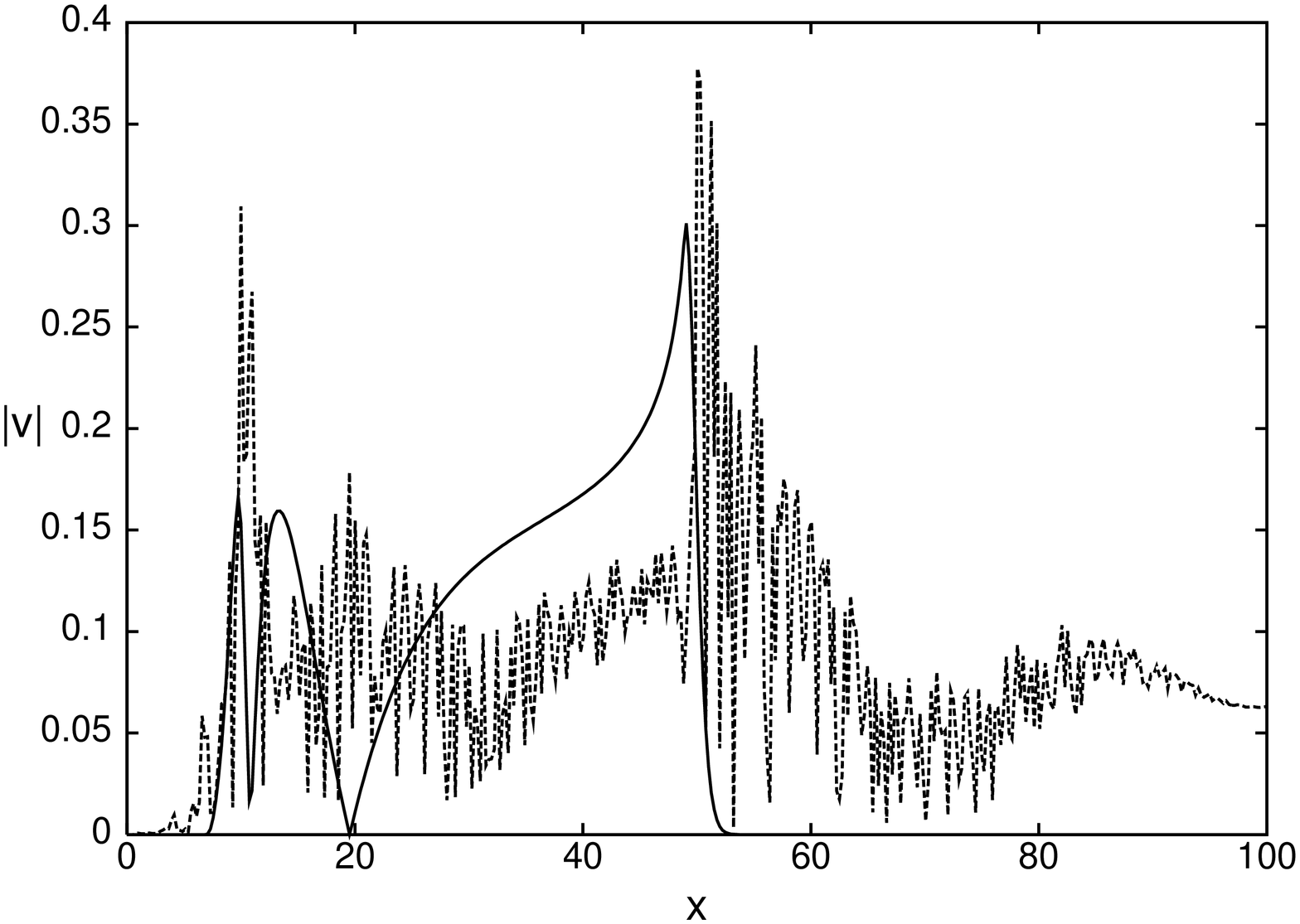,height=10cm,angle=270}}}
\bigskip
\caption{ Evolution of the WKB solution as an initial condition.
              Numerical solution of (\ref{e:epsg}) and (\ref{e:epsl}) at
              $t = 5$: ~--~--~--~; WKB initial condition: ~---~.}
\label{f:wkbcomp}
\end{figure}

To study the problem posed by Eqn.\ (\ref{104}) we begin by
assessing the validity of the WKB analysis.  To this end we use
the WKB eigenfunction of Equations (\ref{aa1}) and (\ref{aa2})
for $n = 2$. This eigenfunction is shown in Fig.\ \ref{f:wkb}.  In
Fig.\ \ref{f:wkbcomp} we show the evolution under the Schr\"odinger 
equations (\ref{e:epsg}) and (\ref{e:epsl}) of this WKB solution as an
initial condition.  It can be seen that a small amount of radiation leaks
to infinity and that the WKB solution is distorted by high frequency
waves.  The high frequency oscillations are due to the non-smooth
WKB initial condition and these high wavenumbers then spread under the
influence of the non-local nature of the governing equations (\ref{e:epsg}) 
and (\ref{e:epsl}).  Ignoring these high frequency waves, the full numerical 
solution compares well with the WKB approximation.  This good comparison
gives confidence in using WKB solutions to explore general initial
value problems.

Taking into account the above considerations, let us consider the
initial value problem for (\ref{e:epsg}) and (\ref{e:epsl}) with a
Gaussian initial condition.  We have two cases: $x_{0}>2M$ and
$x_{0}<2M$. Let us consider first the case $x_{0}>2M$.  In Fig.\ 5
we show the evolution of the initial condition at $t = 5$, at which time
the solution has reached the steady state.  For $M = 5$ this figure shows 
how the initial value is decomposed into the eigenfunction and a small 
amount of probability current that is propagated towards infinity.  This
result can be accounted for by the following WKB analysis.

\begin{figure}
\centerline{
\hbox{\psfig{figure=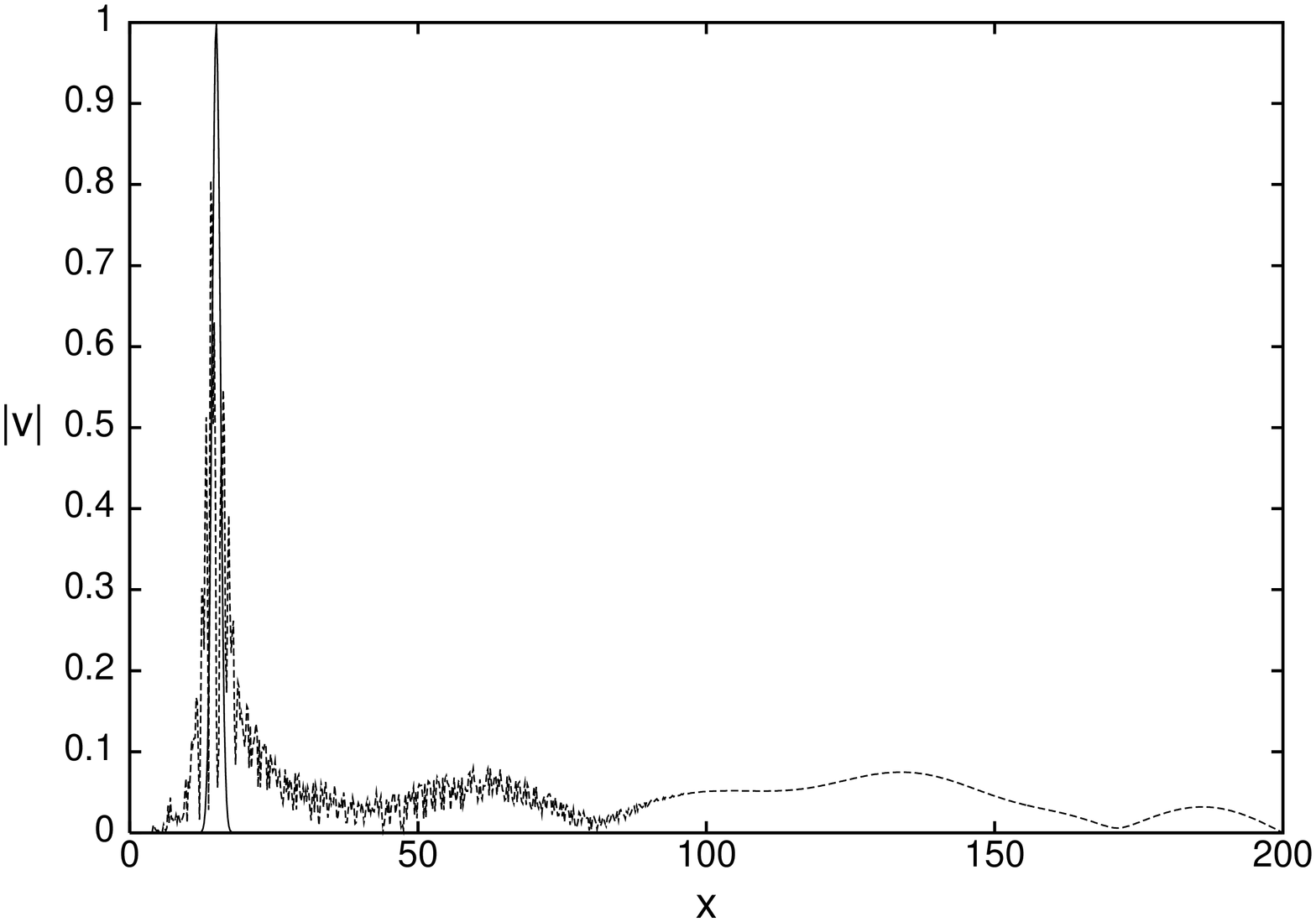,height=8cm,angle=270}}}
\bigskip
\caption{Numerical solution of (\ref{e:epsg}) and (\ref{e:epsl}) for the
       initial condition (\ref{e:betaini}) with $\lambda = 1.0$ and
       $x_{0} = 15.0$ for $M=5$.  Initial condition is outside event horizon at
       $x=10$.  Initial condition: ~---~; solution at $t=5$: ~--~--~--~.
       Solution in $0 < x < 200$.   }
\label{f:m5t5d200}
\end{figure}

\begin{figure}
\centerline{
\hbox{\psfig{figure=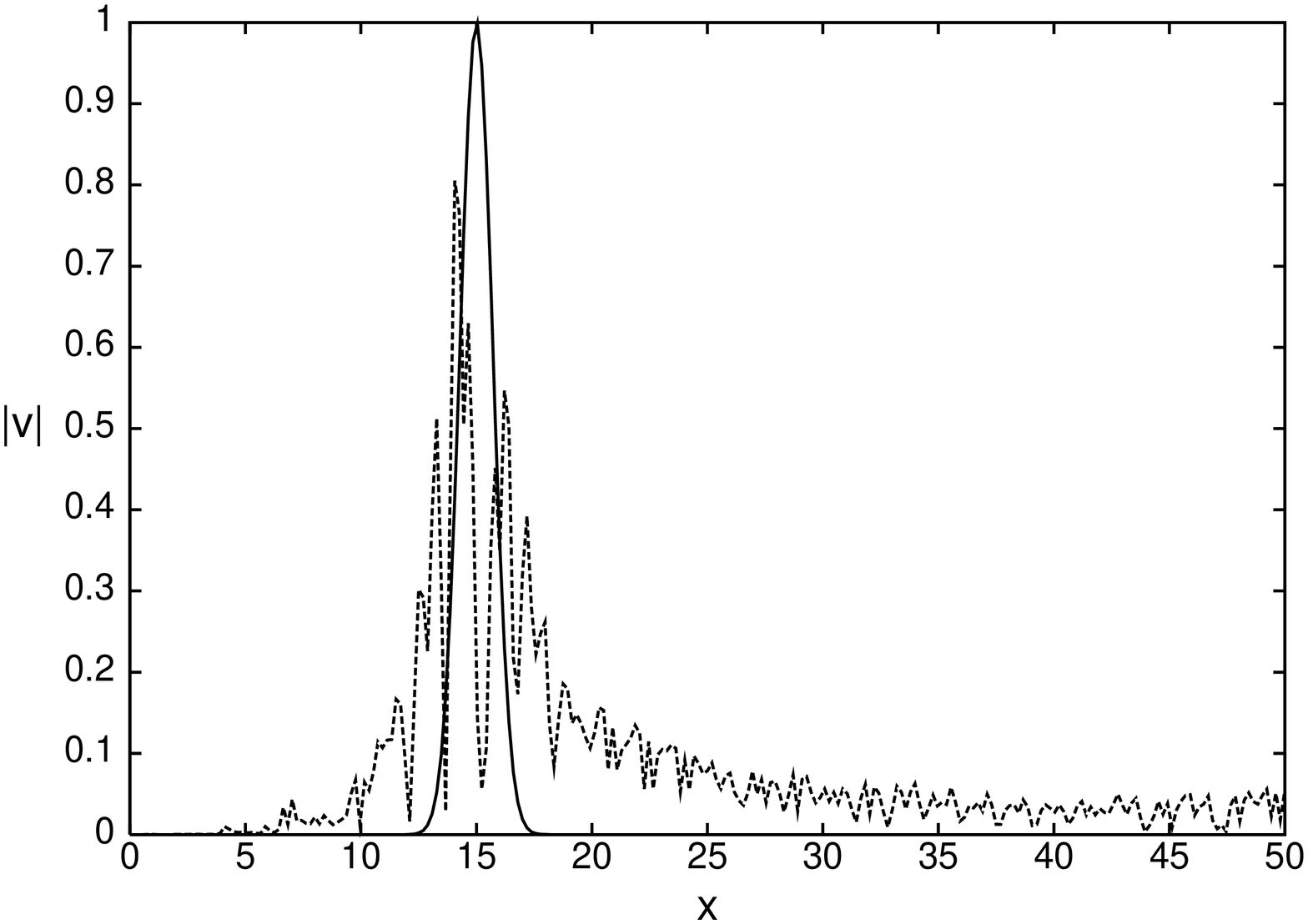,height=8cm,angle=270}}}
\bigskip
\caption{Numerical solution of (\ref{e:epsg}) and (\ref{e:epsl}) for the
       initial condition (\ref{e:betaini}) with $\lambda = 1.0$ and
       $x_{0} = 15.0$ for $M=5$.  Initial condition is outside event horizon at
       $x=10$.  Initial condition: ~---~; solution at $t=5$: ~--~--~--~.
        Solution in $0 < x < 50$ showing decay behind event horizon at
              $x=10$.}
\label{f:m5t5d50}
\end{figure}

The discrete eigenfunctions have their oscillatory part
concentrated around $2M$.  Thus the Gaussian
projected onto these modes decomposes into the oscillations shown
in Figs.~\ref{f:m5t5d200} and \ref{f:m5t5d50}.
The projection onto the continuous spectrum produces
the small current which spreads to infinity with decaying
amplitude.  The high frequency of the radiated waves stems from
the short wave behavior displayed in the WKB solutions.

\begin{figure}
\centerline{
\hbox{\psfig{figure=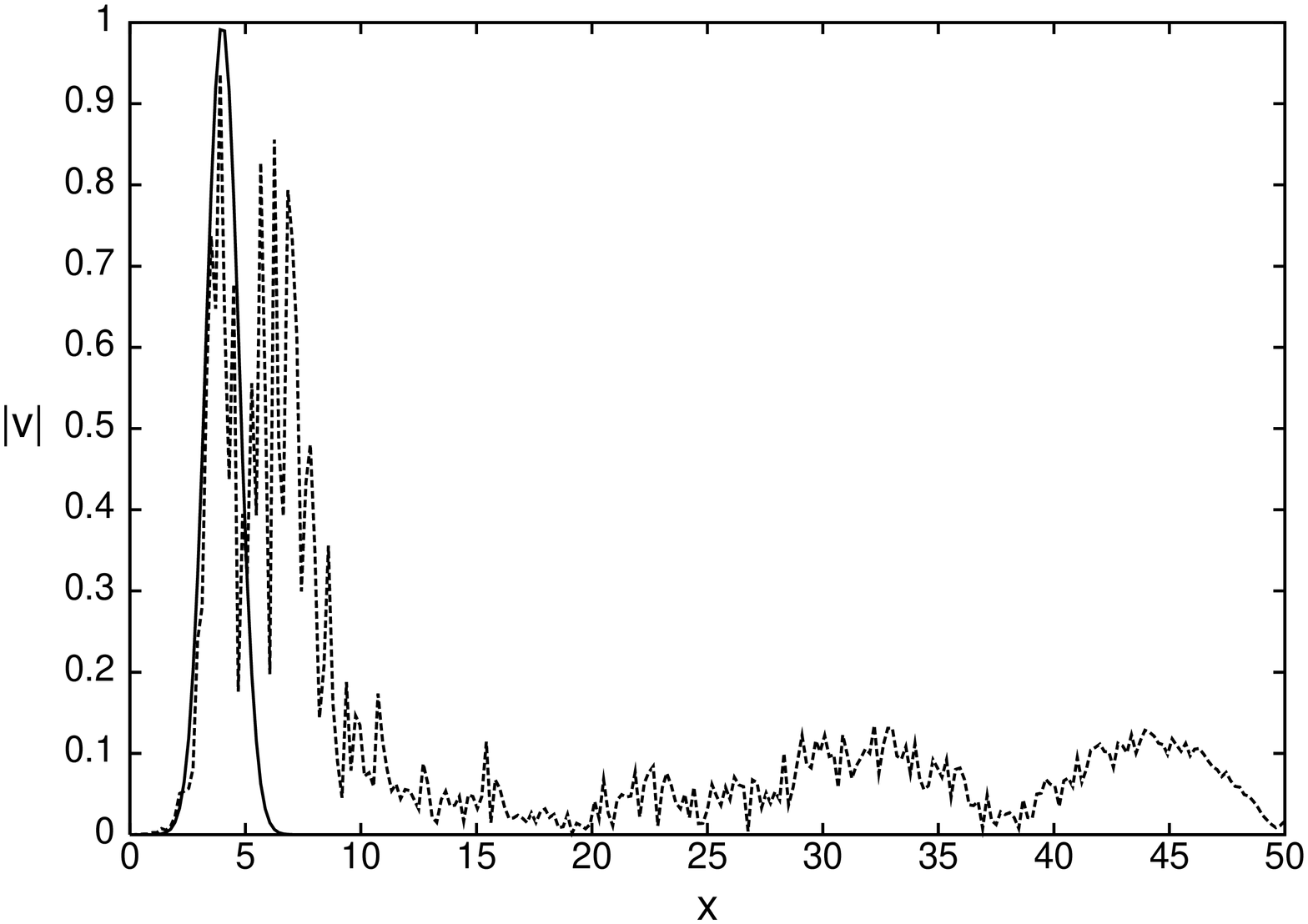,height=10cm,angle=270}}}
\bigskip
\caption{Numerical solution of (\ref{e:epsg}) and (\ref{e:epsl})
        for the initial
       condition (\ref{e:betaini}) with $\lambda = 1.0$ and $x_{0} = 4.0$ for 
       $M=10$.  Initial condition is inside event horizon at $x=20$.  Initial 
       condition: ~---~; solution at $t=10$: ~--~--~--~.}
\label{f:inside}
\end{figure}

The situation for the opposite case $x_{0}<2M$ is shown in 
Fig.~\ref{f:inside}
  The chosen mass parameter was $M = 10$ and the initial value
was taken to be $x_0 = 4$.  Here we have the opposite situation
occurring, namely that the evolution is confined to the left of $x
= 2M$, while for $x_{0}>2M$ we found that the solution was
concentrated to the right of $x = 2M$. This observation may be
explained in terms of the WKB amplitude given in (\ref{23a})
and (\ref{23b}).  From those equations it follows that the
amplitude $A_{2}$ is small at $x = 2M$ and this, in turn, blocks
the passage of the wave function from one region to the other. The
above result can also be understood by noting that at $x = 2M$ the
equations lose the highest order derivative responsible for
propagation and thus the group velocity goes to zero at $x = 2M$
and the probability current also goes to zero (similar to critical
layer absorption at the barrier).

These numerical results confirm our analysis and show that when
the radius of the initial shell is larger than its ``Schwarzschild
radius'', the wave function is concentrated outside $x = 2M$.
While when the radius of the initial shell is inside the
``Schwarzschild radius'', the wave function leaks out very slowly.
Note that the region $x\approx 0$ is always a region of small
probability.  Finally, note that these solutions show the
evolution of the probability density of the radius of the shell
measured as an intrinsic variable.  As they stand, they provide no
information about the space-time geometry generated by the
collapsing shell. This geometry must be calculated and then
quantized making use of the operators generated by the motion of
the shell.  This is the subject of ongoing work.

\section{Conclusions}

We have shown how the quantum evolution of the relativistic collapse of a
shell of matter can be described by making use of the dimensionally
reduced true relativistic Hamiltonian describing this process. In our
analysis we have also given a procedure for quantizing a multi-valued
Hamiltonian, and we have chosen an appropriate lapse function to produce
a tractable Schr\"odinger equation.  This equation has several novel
features which were analyzed in the process of constructing the WKB
solutions.  These solutions have an infinite set of bound states as well as a
set of continuum eigenstates.  The discrete eigenstates were shown to
concentrate away from $R = 0$.  Thus, in analogy to \cite{visser}, this
suggests that the shell does not enter on average the singularity.
The following possibility is also suggested by the present analysis:
an incoming wave packet which classically collapses into the singularity
$R=0$ now, due to the quantum fluctuations, is bounced back but into a new
expanding space-time because of the small quantum fluctuations at $R\sim 0$.
This is a consequence of the Uncertainty Principle which in \cite{visser}
keeps open the throat of the wormhole and in this work bounces the wave
function into a new expanding space-time.

The numerical solutions show that the quantum dynamics of an initial
condition, representing a Gaussian distribution of shell radii, evolves into
a sum over the discrete eigenstates. The critical layer nature of the process
is contrasted with the behavior of other {\it ad hoc} models appearing in
the literature which turns out to be closer to that of the hydrogen atom.

The quantum mechanics developed in this work does not directly answer
questions about the quantum geometry of the space. It does, however, set the
framework for analyzing the local geometry of the space-time around the
shell in terms of the operators defined in this work. In addition, the
viewpoint we have presented provides a completely consistent quantization
from first principles and a choice of time, related to the proper time on
the shell, which explains in WKB terms the behavior of the dynamics
observed in previous {\it ad hoc} models. It must be stressed that the
combination of the WKB approximation and numerics that we used provides
a complete description of a consistent quantum dynamics for the radius of
collapsing shells of matter in a wide parameter regime.

There are several possible directions for future work.  An important study
would be to investigate the external geometry of the system as mentioned above
and the embedding of the shell in this spacetime. This is not an easy job. 
In \cite{louko2},
which discusses the collapse of null shells, it was necessary, in order to
obtain a tractable problem, to define new canonical variables, and the results
for the exterior geometry and embedding were not those that would seem obvious 
at first glance. Finding the exterior geometry would allow us to understand 
whether the quantum mechanical expansion
of the shells we found corresponds to a quantum mechanical tunneling
through a horizon or expands into a different spacetime.  Another direction
would be to study superpositions of states with different values of $M$.
This would lead to a ``fuzzy'' horizon.  Note that in our approximation
(\ref{104}), these superpositions  will have a very different evolution from
the $M$-eigenstate we have used.

Finally, it is important to remark that the results here obtained are 
within the mini-superspace formalism, where one first reduces the classical 
system and then quantizes. Recent results in isotropic (loop) 
quantum cosmology 
indicate that the inverse strategy, namely to first quantize end then reduce,
might yield qualitative different physical predictions at the Planck scale
\cite{Bojowald:2001xe}.

\appendix*

\section{}

We here consider the transition layer behavior at $x = R^*(E)$.  To do
this, expand Eq.\ (\ref{21a}) about $R^*(E)$ to obtain
\begin{equation}
f(x, E) = \frac{-E^2 /2 + x^2 - Mx}{x^{3/2} (x - 2M)^{1/2}} \approx 1 + \frac
{\theta^{\prime 2}}{2},
\end{equation}
or
\begin{equation}
1 + f^{\prime} (R^*, E) [x - R^*(E)] = 1 + \frac {\theta^{\prime 2}}{2}.
\end{equation}
Now $\theta^{\prime 2}$ is replaced by the second derivative of the boundary
layer function, $w$, and we obtain
\begin{equation}
w^{\prime \prime} + f^{\prime} [R^*(E) , E) [x - R^*(E)]w.
\end{equation}
Since $f^{\prime} < 0$, we have the Airy equation for $w$ \cite{tim}.  It has
solutions of the form
\begin{equation}
w = {\rm Ai}\, (\beta^{1/3} [x - R^*(E)]),
\end{equation}
\begin{equation}
\beta = 2 \left| \frac{-\left( M^{2} + E^{2} \right) R^{*} + \frac{3}{2}
ME^{2}}{R^{*2} - 2MR^{*}} \right| .
\end{equation}
The Airy function matches the oscillatory solution as $x - R^*(E) \rightarrow
- \infty$, and decays exponentially as $x - R^*(E) \rightarrow \infty$.
In the matching of the phases of the Airy function and $\sin \varphi_2 (x)$,
we obtain the WKB energy quantization rule in the usual way \cite{tim}.

To consider the matching at the point $\tilde R(E)$, we follow the same
procedure.  Expanding around $\tilde R(E)$, we obtain
\begin{equation}
f^{\prime} [\tilde R(E)] [x - \tilde R(E)]u = \frac{1}{2\pi}
\int^{\infty}_{-\infty} |\sinh k| e^{ikx} \hat u \: dk
\end{equation}
Changing variables to $x^{\prime} \equiv x - \tilde R(E)$, we have
\begin{equation}
f^{\prime} [\tilde R(E)] x^{\prime} u = \frac{1}{2\pi}
\int^{\infty}_{-\infty}|\sinh k| e^{ikx^{\prime}}\hat u \: dk.
\end{equation}
This equation can be solved by means of Fourier transforms to give
\begin{equation}
i\frac{d}{dk} \hat u = |\sinh k| \hat u,
\end{equation}
which yields
\begin{equation}
\hat u = A e^{-i \int^k_0 |\sinh \xi |d\xi},
\end{equation}
or
\begin{equation}
\hat u =
\cases{ A e^{-i(\cosh k - 1)}, & $k \geq 0$, \cr
                   A e^{-i(\cosh k + 1)}, & $k < 0$. \cr }
\end{equation}

Note that this function is not analytic at $k = 0$.  Thus, the only path of
integration for the inversion is the real axis.  We obtain
\begin{equation}
u(x^{\prime}) = \frac{A}{2\pi} \int^{\infty}_0 e^{ikx^{\prime}} e^{-i(\cosh k
- 1)}dk
+ \frac{A}{2\pi} \int^0_{-\infty} e^{ikx^{\prime}} e^{i(\cosh k - 1)}dk.
\end{equation}
For $x > 0$ the points of stationary phase satisfy
\begin{equation}
|\sinh k| = x^{\prime},
\end{equation}
as they should to match the behavior for $x > 0$.  On the other hand,
for $x < 0$ there are no points of stationary phase.  The integral,
after an integration by parts, shows that $u$ decays as
\begin{equation}
u(x) \sim \frac{{\rm constant}}{x^{\prime}} \qquad {\rm as}\;\; x^{\prime}
\rightarrow -\infty.
\end{equation}
This is due to the nonanaliticity of $|\sinh k|$.

This behavior shows us the reason for the sharp decay of amplitude observed
in the numerics.  Note that although this analysis is only local around $x
= \tilde R(E)$, it is sufficient to complete the WKB solution and to
explain the numerics.

\begin{acknowledgments}

We would like to thank C. Beetle, J. Guven,  and B. Whiting
for discussions, and specially  Karel 
Kucha\v r for many
fruitful discussions during the various stages of this work.
This work was supported in part by UNAM-DGAPA Project No. IN 106097 and by
CONACyT Project No. G  25427-E and J32754-E. Our special thanks to 
Ana Guzman for her help in computations.

\end{acknowledgments}

\end{document}